# Temporal data series of COVID-19 epidemics in the USA, Asia and Europe suggests a selective sweep of SARS-CoV-2 Spike D614G variant

Taimá N. Furuyama*[1], Fernando Antoneli*[3], Isabel M. V. G. Carvalho[2], Marcelo R. S. Briones[2], Luiz M. R. Janini[4]

[1] Laboratório de Retrovirologia e Departamento de Medicina, Escola Paulista de Medicina, Universidade Federal de São Paulo, São Paulo Brazil.

[2] Laboratório de Genômica Médica e Departamento de Informática em Saúde, Escola Paulista de Medicina, Universidade Federal de São Paulo, São Paulo Brazil.

[3] Laboratório de Parasitologia, Instituto Butantan, São Paulo, Brazil.

[4] Laboratório de Retrovirologia e Departamento de Microbiologia, Imunologia, Parasitologia, Universidade Federal de São Paulo, São Paulo, Brazil.

**Both share first author status.*

**Correspondence**:

Fernando Antoneli (fernando.antoneli@unifesp.br);
Marcelo R. S. Briones (marcelo.briones@unifesp.br)
Rua Pedro de Toledo 669, L4E
São Paulo, SP, 04039-032, Brazil.
Tel: 011 5576-4537

**Keywords:** COVID-19, SARS-CoV-2, D614G, mutation, population genetics, coronavirus.

**ABSTRACT**

The COVID-19 pandemic started in Wuhan, China, and caused the worldwide spread of the RNA virus SARS-CoV-2, the causative agent of COVID-19. Because of its mutational rate, wide geographical distribution, and host response variance this coronavirus is currently evolving into an array of strains with increasing genetic diversity. Most variants apparently have neutral effects for disease spread and symptoms severity. However, in the viral Spike protein, which is responsible for host cell attachment and invasion, an emergent variant, containing the amino acid substitution D to G in position 614 (D614G), was suggested to increase viral infection capability. To test whether this variant has epidemiological impact, the temporal distributions of the SARS-CoV-2 samples bearing D or G at position 614 were compared in the USA, Asia and Europe. The epidemiological curves were compared at early and late epidemic stages. At early stages, where containment measures were still not fully implemented, the viral variants are supposed to be unconstrained and its growth curves might approximate the free viral dynamics. Our analysis shows that the D614G prevalence and the growth rates of COVID-19 epidemic curves are correlated in the USA, Asia and Europe. Our results suggest a selective sweep that can be explained, at least in part, by a propagation advantage of this variant, in other words, that the molecular level effects of D614G have sufficient impact on population transmission dynamics as to be detected by differences in rate coefficients of epidemic growth curves.

**INTRODUCTION**

The Coronavirus Disease 2019 (COVID-19) outbreak is caused by SARS-CoV-2 (Severe Acute Respiratory Syndrome Coronavirus type 2) and was declared by the World Health Organization (WHO) as a pandemic in March 11, 2020 (World Health Organization, 2020c, 2020d). As of June 15, 2020, the virus has already infected more than 7.3 million people and caused more than 400,000 deaths worldwide (5.7% fatality ratio) (Dong et al., 2020; Johns Hopkins University, 2020). This is the third coronavirus caused outbreak in less than 20 years (World Health Organization, 2020a). From November 2002 to May 2004, SARS-CoV-1 (Severe Acute Respiratory Syndrome caused by Coronavirus type 1) affected 26 countries worldwide, accounted 8,096 confirmed cases and 774 deaths (9.6% fatality ratio) (Drosten et al., 2003; Ksiazek et al., 2003; Lee et al., 2003; Peiris et al., 2003; Zhong et al., 2003; Centers for Disease Control and Prevention - Department of Health and Human Services, 2004; World Health Organization, 2004; Centers for Disease Control and Prevention, 2017). MERS-CoV (Middle East Respiratory Syndrome caused by Coronavirus) spread to 27 countries around the globe, totalizing 2,519 confirmed cases and 866 deaths (34.4% fatality ratio) continuously since April 2012 (Zaki et al., 2012; Hijawi et al., 2013; Centers for Disease Control and Prevention, 2019; World Health Organization, 2019, 2020b). Several conditions contribute to the transmission speed of SARS-CoV-2, such as transmission during the asymptomatic phase and wide human susceptibility to this pathogen (Arons et al., 2020; Fam et al., 2020; Gandhi et al., 2020). The central concern for governments and general population is the collapse of healthcare systems and lack of essential care. Therefore, cumulative information about the SARS-CoV-2 mechanisms inside the host cell, its epidemiology and its genetic patterns are necessary to halt the virus spread, to prevent the disease and heal the infected individuals.

The comparison of several SARS-CoV-2 strains with the Wuhan reference genome genome (GenBank accession NC_045512) reveals a G to A transition at position 23,403 that leads to a D to G amino acid substitution at position 614 in the Spike protein. Molecular evidence suggests that this substitution is advantageous for viral propagation *in vitro* because of increased Spike protein abundance and reduced shedding (Zhang et al., 2020). A previous study by Korber and collaborators (Korber et al., 2020) conjectures that the D614G substitution in SARS-CoV-2 Spike protein could be responsible for higher transmission rates observed in a global scale. The study shows that there is a higher prevalence of D614 in China and in the United States before March 2020 while after March 2020 the G614 prevalence significantly increases in Europe and United States. If there is a correlation between the D614G variant prevalence and higher SARS-CoV-2 transmission, then the epidemiological data might reveal a significant correlation between D614G prevalence and the growth rate coefficients of epidemic curves globally. Here we present evidence that the prevalence of D614G is correlated with increased growth rate coefficients in temporal series of COVID-19 epidemiological data. The relative dynamics of D614 and G614 variants observed is what would be expected in the case of a selective sweep.

**METHODS**

**D614G variant data**
Data on prevalence of D614 or G614 was obtained from the Los Alamos distribution map of D614 and G614 SARS-CoV-2 (Elbe and Buckland-Merrett, 2017; Shu and McCauley, 2017; Los Alamos National Laboratory, 2020). The data was downloaded as "data-2020-05-20.csv" file for "all" time range. The time range considered was 21 days (3 weeks). Data: Obtained from Coronavirus COVID-19 Global Cases by the Center for Systems Science and Engineering (CSSE) at Johns Hopkins University; the Red Cross; the Census American Community Survey; the Bureau of Labor and Statistics:
(https://github.com/CSSEGISandData/COVID-19).

**Epidemic data**
The epidemic growth rates were obtained from Coronavirus COVID-19 Global Cases by the Center for Systems Science and Engineering (CSSE) at Johns Hopkins University; the Red Cross; the Census American Community Survey; and the Bureau of Labor and Statistics data (Johns Hopkins University, 2020; American Red Cross, 2020; United States Census Bureau, 2020; U.S. Bureau of Labor Statistics, 2020). The time range considered was the same as the considered in the Los Alamos data, being the First day of time series of confirmed cases on February 18, 2020 and the last day of time series of confirmed cases in May 27, 2020.

**Countries, States, Counties and time ranges**
Six states and 325 counties of the USA were considered in the analysis further grouped as East or the West Coast. Accordingly, West Coast States data included (with number of counties in parentheses): Oregon (34), Washington (39), California (57), being a total 129 of the West Coast. The East Coast States included (number of counties in parentheses): New York (58), Connecticut (9), Virginia (130) totaling 196 counties of the West Coast. The time range was divided into an early start period and a late start period. The early start considered counties such that the epidemics started in the range 2/18/20 – 3/18/20 (98 counties). The late start considered counties such that the epidemics started in the range 3/18/20 – 5/27/20 (325 counties). Nine counties did not have at least 21 days of non-zero time series. The analysis was also made for other regions of the world, considering a broader division: World (some countries have more than 1 administrative region): Western Countries (Europe): Belgium (1), Denmark (3), France (9), Germany (1), Italy (1), Luxembourg (1), Netherlands (5), Portugal (1), Spain (1), United Kingdom (11); total number of administrative regions = 34. Eastern Countries (Eastern Asia and Oceania): Australia (8), Bangladesh (1), China (32), India (1), Japan (1), South Korea (1), Singapore (1), Taiwan (1), Thailand (1), Vietnam (1); total number of administrative regions = 48. The first day of time series of confirmed cases was 1/22/20 (eastern countries); 2/15/20 (western countries) and the last day of time series of confirmed cases was 5/27/20.

Regarding the early and late epidemic periods, the time frame was: Early start: countries such that the epidemics started in the range: 2/15/20 – 3/15/20 (western countries; 19 regions) and 1/22/20 – 5/27/20 (eastern countries; 47 regions). Late start: countries such that the epidemics started in the range: 3/15/20 – 5/27/20 (western countries;15 regions) and 1/22/20 – 5/27/20 (eastern countries; 47 regions). One region did not have at least 21 days of non-zero time series and the eastern regions are all early starters, so in this case we compared the eastern regions with the early western region and the late western regions.

**Logistic models**
The logistic model parameters were obtained from logistic regression (Spiegelhalter, 1986) using python 3 with pandas libraries (https://pandas.pydata.org/) and scikit-learn (https://scikit-learn.org/stable/about.html). Plots were generated with matplotlub (https://matplotlib.org/) and seaborn (https://seaborn.pydata.org/index.html).

**RESULTS**

The initial analysis consisted in the logistic models derived from US data comparing US East coast (predominantly G614) with West coast (predominantly D614), the Asia-Europe data in West (predominantly G614) and East (predominantly D614) in the early and late epidemic stages. **Figure 1** depicts the plots of logistic models with the corresponding logistic model (blue line) and its confidence band (light blue shading). **Figure 1A** shows the Early Start Counties of USA, **Figure 1B** shows the Late Start counties of USA, **Figure 1C** shows the Early Start countries of the Asia-Europe axis and **Figure 1D** shows the Late Start countries of Asia-Europe axis. The comparisons between the logistic models in early and late epidemic stages show that at the early stages the growth rates between West and East, either US or Europe, are significantly different, while in late stages the West-East differences are not significant. This test therefore suggests that in the early epidemic stages the predominant variant pattern reflected a "founder effect", especially in the US, where the West coast infections derived from an Asian D614 type whereas in the East cost the infection dynamics started with European derived ancestors, of the G614 type. This is observed from D614G distribution data. The early epidemic stage in both US and Asia-Europe show significant differences in the odds ratios in the West and East portions, showing that the growth rates might be impacted by the G614 substitution. At the late stages the growth rates are not distinguishable. European late stages show odds ratios < 1 and in the US the odds ratios drop from 1.16 to 1.03.

The growth curves of the D614 and G614 variants in West US, East US, West Asia-Europe and East Asia-Europe at different time periods, of 10 days each, are shown in **Figure 2.** The time series of frequency variants (D/G) reveal that irrespective of geographic region and early and late epidemic stages, the G614 variant increases and surpasses the D614. **Figure 2A** shows the USA dynamics and **Figure 2B** shows the Asia-Europe dynamics. Each curve corresponds to the variant frequency in the corresponding geographic region. The frequencies of variants in the same region are complementary to each other. In US East the G614 started at approximately 80% and D614 at ~20% with subsequent increase of G614 to 100% and extinction of D614. In US West, G614 started at ~1% and D614 ~99% and after 70 days G614 increased to more than 60% whereas D614 decreased to less than 40% (**Figure 2A**). This type of dynamics is highly suggestive of a selective sweep because of an increased infectivity/replication rate of G614. The effect is very similar in Europe West and East where irrespective of the initial frequencies of D614 and G614, the later always predominates after 60 days (**Figure 2B**).

In **Supplementary Table S1** the frequencies of D614 and G614 are compared in the East and West coast for the US while **Supplementary Table S2** depicts the Asia-Europe frequencies of D614 and G614. This data show that the growth rate of G614 is significantly higher than the D614 growth rate. Also, it indicates that the phenomenon is global, not restricted to a geographic location of specific host population. The initial cases in the East coast are likely to have originated from European strains (predominantly G614) whereas in the West coast the initial infections were caused by Asian strains D614 predominant in that Continent at that time.

**DISCUSSION**

The first confirmed COVID-19 case in the United States was in the state of Washington on January 20, 2020 (Centers for Disease Control and Prevention, 2020; Holshue et al., 2020). This would explain the similarities in transmission processes in the US West coast when compared to China, Japan and Taiwan. On the other hand, in the US East coast, especially New York, the likelihood of the beginning of the COVID-19 epidemic is of European origin.

In the present work we analyzed the frequencies of D614 and G614 variants in US West and East and Asia-Europe West and East. We have shown that irrespective of initial frequencies of these variants at early epidemic stages, G614 always predominates, and very quickly either becomes fixed or significantly surpasses D614 after 60 days of the initial infection. In a previous study (Korber et al., 2020) conjecture on founder effects and selection when D614 and G614 are compared. The analysis shown in **Figure 1** indicates that at early epidemics the founder effect is more prominent and at later stages the selection is more significant.

The populational dynamics depicted in **Figure 2**, indicates a selective sweep of G614 over D614. A selective sweep is a population genetics process in which a novel beneficial mutation increases its frequency to a point where it reaches 100% and is therefore, "fixed" in the population (Hermisson and Pennings, 2005). The current COVID-19 epidemics and the discovery of a Spike protein variant with a mutation from D to G at position 614 has given an opportunity to show such process in action due to the detailed epidemiological data and viral genome sequence availability. Our analysis also shows that the founder effect and selective sweep are not specific to a country or region, which suggests that the selective advantage of G614 over D614 is global and occurs irrespective of the genetic variation and ethnic background of the host populations.

Although the results indicate that there is a robust difference between the D614G variant and the epidemic growth rate curve it is important to point there are several mutations occurring in the viral genome, which could compose a mutation balance in the viral fitness. That is, it is unlikely that the fitness increase by G614 alone drive the epidemic curves. As shown by (Korber et al., 2020) other mutations hitchhike around G614 by recombination and therefore the combined fitness of several mutations increase the fitness of SARS-CoV-2 to a point that explains the selective sweep observed in **Figure 2**. Nevertheless, we provide robust population evidence for the hypotheses raised by which combine founder effect and selection and believe that G614 predominance over D614 is an example of selective sweep in a viral population.

**Author contributions**
IMVGC, LMRJ, MRSB, FA: Data analysis planning and conceptualization. TNF, FA: Performing the data analysis. All authors: writing and editing the manuscript.

**Statement of conflict of interests**
The authors declare no conflict of interests.

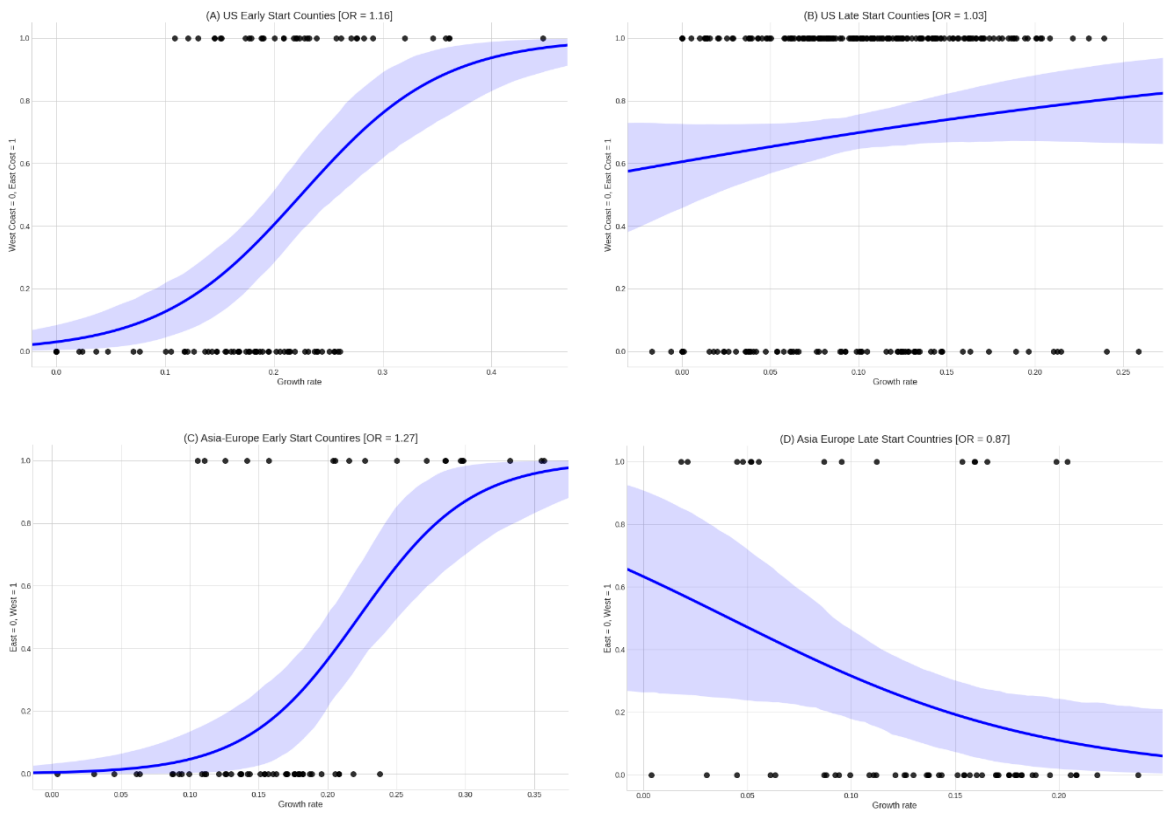

**FIGURE 1.** Plots of logistic models. Panel **(A)** shows the Early Start Counties of USA, panel **(B)** shows the Late Start counties of USA, panel **(C)** shows the Early Start countries of the Asia-Europe axis and panel **(D)** shows the Late Start countries of Asia-Europe axis (black dots). The blue curve is the corresponding logistic model with its confidence band (light blue shading). In each panel, the horizontal axis is the growth rate of the initial segment of 21 days of the corresponding time series of confirmed cases, and the vertical axis is a binary variable indicating the corresponding region where the time series is from (East/west coast county in USA, panels **(A)** and **(B)** or East/West country in Asia-Europe axis, panels **(C)** and **(D)**).

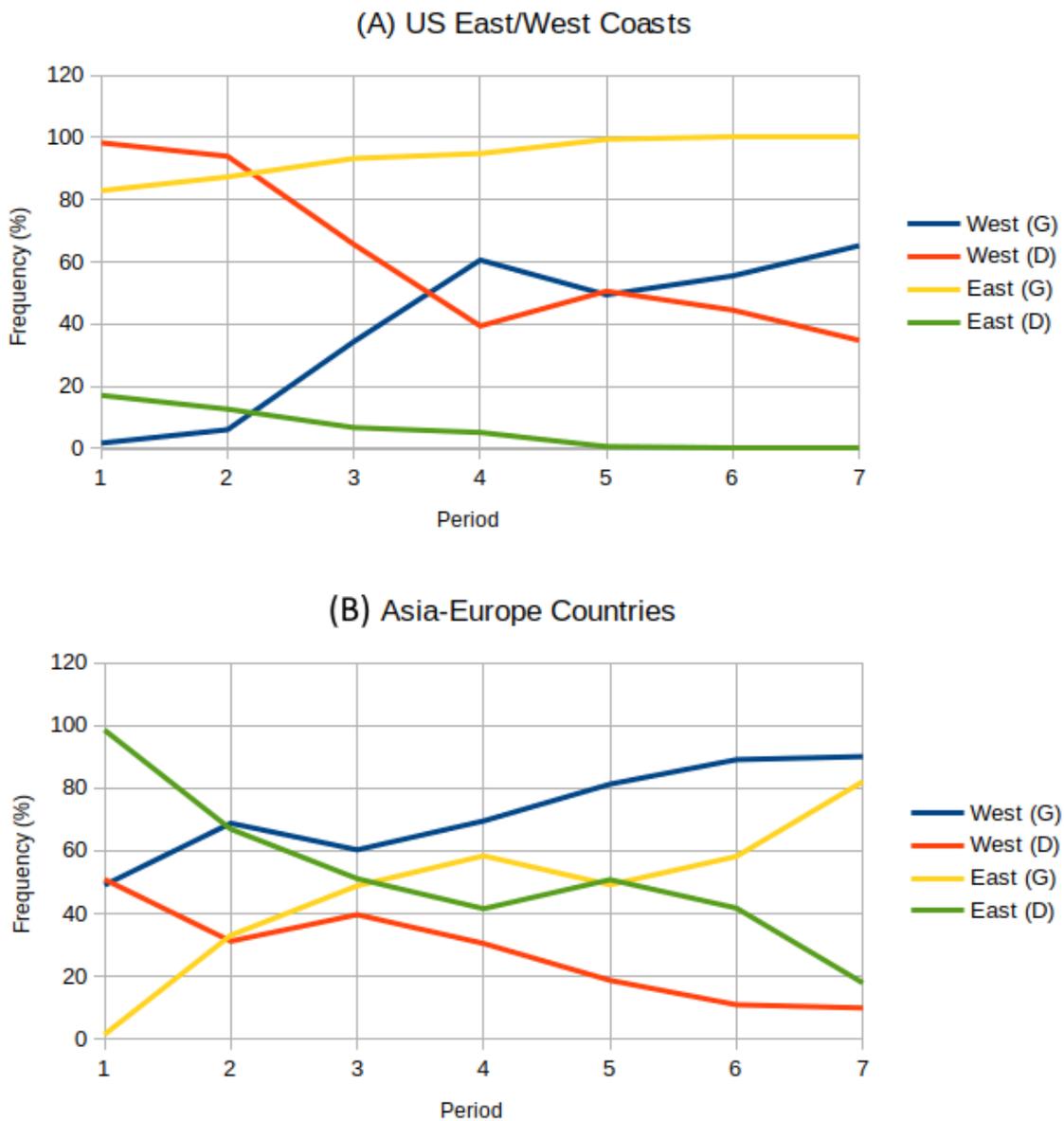

**FIGURE 2.** Time evolution of the frequency viral variants (D/G) with respect to the residue 614 of Spike Protein. Panel **(A)** shows the USA and panel **(B)** shows the Asia-Europe axis. In each panel, the horizontal axis shows time period of time (corresponding to 10 days each) and the vertical axis show the frequency. Each curve corresponds to the variant frequency in the corresponding region, according to the side legends. The frequencies of variants in the same region are complementary to each other. Adapted from (https://cov.lanl.gov/apps/covid-19/map/) (Korber et al., 2020).

**Supplementary Table S1 |** The frequencies of SARS-Cov2 D614 and G614 variants in the East and West coasts of the US.

| | US Data | | | | | |
|---|---|---|---|---|---|---|
| **Period** | **West Coast** | | | **East Coast** | | |
| | G614 | D614 | Total | G614 | D614 | Total |
| 1 | 1 | 56 | 57 | 0 | 0 | 0 |
| 2 | 12 | 186 | 198 | 63 | 13 | 76 |
| 3 | 183 | 349 | 532 | 537 | 78 | 615 |
| 4 | 320 | 208 | 528 | 304 | 22 | 326 |
| 5 | 211 | 216 | 427 | 402 | 22 | 424 |
| 6 | 111 | 89 | 200 | 156 | 1 | 157 |
| 7 | 60 | 32 | 92 | 41 | 0 | 41 |

\* Data on the prevalence of variant of the virus with respect to the residue 614 of Spike Protein is prevalent in the infected population (D or G). Data from the site "Distribution of D614 and G614" (https://cov.lanl.gov/apps/covid-19/map/) from (Korber et al., 2020).

**Supplementary Table S2 |** The frequencies of SARS-Cov2 D614 and G614 variants in the East and West in the Asia-Europe data.

| | Asia-Europe Data | | | | | |
|---|---|---|---|---|---|---|
| **Period** | **West** | | | **East** | | |
| | G614 | D614 | Total | G614 | D614 | Total |
| 1 | 53 | 55 | 108 | 11 | 767 | 778 |
| 2 | 574 | 260 | 834 | 41 | 83 | 124 |
| 3 | 1152 | 757 | 1909 | 259 | 271 | 530 |
| 4 | 2986 | 1299 | 4267 | 353 | 251 | 604 |
| 5 | 4038 | 929 | 4967 | 151 | 156 | 307 |
| 6 | 1946 | 238 | 2184 | 67 | 48 | 115 |
| 7 | 892 | 98 | 900 | 69 | 15 | 84 |

\* Data on the prevalence of variant of the virus with respect to the residue 614 of Spike Protein is prevalent in the infected population (D or G). Data from the site "Distribution of D614 and G614" (https://cov.lanl.gov/apps/covid-19/map/) from (Korber et al., 2020).